\documentstyle[12pt]{article}
 
\begin{document}

\author{\and F.~Haas \& J.~Goedert \\
Instituto de F\'{\i}sica, UFRGS\\
Caixa Postal 15051\\
91500-970 Porto Alegre, RS - Brazil}
\title{On the Hamiltonian Structure of Ermakov Systems}
\date{\relax}
\maketitle

\vfill

\vskip 1cm\noindent
{\bf Short title:} Hamiltonian Ermakov Systems.

\newpage

\centerline{\bf Abstract}

\vskip 2cm\noindent
A canonical Hamiltonian formalism is derived for a class of 
Ermakov systems specified by several different frequency 
functions.  
This class of systems comprises all known  cases of Hamiltonian 
Ermakov systems and can always be reduced to quadratures. The 
Hamiltonian structure is explored to find exact solutions for the 
Calogero system and for a noncentral potential with dynamic 
symmetry.  Some generalizations of these systems possessing exact 
solutions are also identified and solved.  

\vskip 2cm
 
\section{Introduction}
Ermakov systems have been intensively studied since the late 
sixties both in view of their nice mathematical properties and of 
the application potential of their celebrated invariants. More 
recently, the identification of additional structures in the 
original Ermakov-Pinney system and in its generalized version, the 
Ermakov-Lewis-Ray-Reid (ELRR) system (see e.~g.~\cite{gov-leach-%
pla}~and~ \cite{rog-ray} and references therein for an updated 
appraisal of the subject and its applications) has called for 
extra attention. After the early scrutiny of the symmetry 
properties of the ELRR system \cite{gov-leach-pla} and
the exploration of several 
generalization schemes \cite{ath,goatle}, the attention 
has been, more recently, centered on the joint existence of a 
second independent constant of motion or a Hamiltonian for 
sub classes of Ermakov systems \cite{goed}--\cite{nturo}.  This 
particular topic is important {\it per se} because, for Ermakov 
systems, the existence of a second constant of motion usually 
implies, as shown in section $2$, complete integration.
Most importantly, however, the Hamiltonian structure is 
fundamental in various contexts of Physics like quantization and 
perturbation theory.  

A generic ELRR system for two independent variables is given
by equations  (\ref{ermaori1}--\ref{ermaori2}) below.
This paper considers Ermakov systems for which $\omega$ is 
generalized to be a function of the variables $x$ and $y$ besides 
the time $t$. By choosing the function $f$ and $g$ appropriately, 
it is possible to fit the equations into a Hamiltonian formalism and 
eventually derive another constant of motion, distinct from the 
commonly known Lewis-Ray-Reid invariant (LRRI) given by equation 
(\ref{invaori}).  When two integrals exist for this Hamiltonian 
Ermakov system then we show that the equations are completely 
integrable.  

A Hamiltonian structure for the ELRR system can be enforced by 
specializing the arbitrary functions $f$, $g$ and $\omega$.  For 
$w=\omega(t)$, that is, when $\omega$ is a function of $t$ only, 
the Hamiltonian constraint involves only $f$ and $g$. This 
specialization has already served to important applications 
\cite{goncha} where the potential in the resulting Hamiltonian is 
essentially built with what remains of the functions $f$ and $g$.  
In a more general situation where $\omega$ depends also on the 
dynamic variables, the Hamiltonian constraint results less 
restrictive and applies for a wider class of admissible systems.  

In this paper we consider the Hamiltonian property of the 
generalized ELRR system in which $\omega$ may depend not only on 
time but also on $x$ and $y$. In section $2$, a quadratic form in 
the momenta is proposed for the Hamiltonian and the consequences 
of this choice are explored analytically. In this way, all known 
cases of Hamiltonian Ermakov systems in two spatial components are 
recovered and generalized. The general class of nonlinear systems 
thus determined is, in addition, shown to be exactly integrable.  
This remarkable fact encounters application in several areas of 
Physics. In section $3$ we apply the technique in two different 
situations which in our understanding illustrate its applicability 
both in recovering results already known in the literature and in 
identifying new exactly solvable models. Among the new results we 
quote a modified version of the Calogero potential \cite{Calo} and 
a variation of a noncentral Hartmann potential, known to possess 
dynamic symmetry \cite{winternitz}--\cite{boschi}.  

The most general system in two configuration variables that
qualifies for an Ermakov or ELRR system is usually written as 
\begin{eqnarray}
\label{ermaori1}
\ddot{x} +\omega^2 x &=& \frac{1}{yx^2}f(y/x) \,,\\
\label{ermaori2}
\ddot{y} +\omega^2 y &=& \frac{1}{xy^2}g(x/y)\,, 
\end{eqnarray}
where the over dot stands for time derivative, $f$ and $g$ are 
arbitrary functions of their arguments and $\omega$ is an arbitrary 
function of time $t$ and of $x$, $y$ and their time derivatives of 
first and higher orders. For practical reasons, we shell consider 
only $\omega=\omega(x,y,t)$.  

The system of equations (\ref{ermaori1}--\ref{ermaori2}) 
possesses the Lewis-Ray-Reid invariant \cite{ray}
\begin{equation}
\label{invaori}
I=\frac{1}{2}(x\dot{y}-y\dot{x})^2 + \int^{y/x}f(s)\,ds + 
\int^{x/y}g(s)\,ds\,.
\end{equation}

The invariant (\ref{invaori}) persists for arbitrary dependence of 
$\omega$ on $x$ and $y$ \cite{gov-leach-pla,goed}. 
We can therefore  merge $\omega^2$ and $g$ in a single function 
and redefine $f$ according to the following rules:
\begin{eqnarray}
\label{Omeg}
\omega^2 &\longmapsto& \Omega^2 \equiv \omega^2(x,y,t) -
\frac{1}{x
y^3}\,g(x/y) \,,\\
\label{efe}
f(s) &\longmapsto& F(s) = f(s) - \frac{1}{s^2} g(1/s)\,,\\
\label{ge}
g(s) &\longmapsto& 0\,.
\end{eqnarray}
These redefinitions simplify future considerations and cast 
the ELRR system and the LRRI in the more compact forms
\begin{eqnarray}
\label{erma1}
\ddot{x} +\Omega^2 x &=& \frac{1}{yx^2}F(y/x)\,, \\
\label{erma2}
\ddot{y} +\Omega^2 y &=& 0\,, \\
\label{inva}
I=\frac{1}{2}(x\dot{y}\!\!&-&\!\!y\dot{x})^2 + 
\int^{y/x}F(s)\,ds\,.
\end{eqnarray}

The transformations (\ref{Omeg}--\ref{ge}) also indicate that the 
conventional ELRR system comprises only {\it two} arbitrary 
functions and not three as implied by the traditional notation.  
Of course, all previous results found in the literature, obtained 
in the standard notation, remain true. Notice, however, that we can 
always consider the Hamiltonian property of ELRR system in the 
form (\ref{erma1}--\ref{erma2}) without any loss of generality.  

\section{Hamiltonian Formalism}

As already mentioned, $\omega$ in (\ref{ermaori1}--\ref{ermaori2}) 
can be a function of time and of any combinations of the dynamic 
variables $x$, $y$ and their time derivatives of arbitrary order.  
In this work we shell consider the Hamiltonian property of the 
ELRR system for which the frequency function is allowed to depend 
not only on time (as usual) but also on $x$ and $y$.  In 
this case, the resulting constraint becomes less restrictive and 
we can find a much wider class of ELRR systems satisfying 
the Hamiltonian property.  

We shall consider for the Hamiltonian of the Ermakov system 
(\ref{erma1}--\ref{erma2}) the  function
\begin{equation}
\label{H}
H = \frac{A}{2}p_{x}^2 + Bp_{x}p_{y} + \frac{C}{2}p_{y}^2 +
V(x,y,t)\,,
\end{equation}
where $A$, $B$ and $C$ are numbers such that $AC - B^2 \neq 0$ and 
$V(x,y,t)$ is a potential function depending on time and on the 
spatial variables.  

The {\it ansatz} (\ref{H}) is justified in Douglas theory for two 
dimensional Lagrangian systems \cite{douglas}, where it is shown 
that the coefficients of the quadratic terms in the velocities in 
a Lagrangian are constants of motion, at least for velocity-free 
force fields. In particular, these coefficients can be taken as 
numerical constants. As one can show,  the addition of a term 
linear in the momenta does not alter the generality of the 
description. Finally, two cases of Hamiltonian Ermakov systems 
known in the literature are of this
proposed form \cite{goed,lejar}.  

We now  impose that the canonical Hamilton equations   
generate the ELRR system (\ref{erma1}--\ref{erma2}).  
The Hamiltonian function (\ref{H}) generates the
following equations for the motion of the system: 
\begin{eqnarray}
\dot x &=& Ap_x + Bp_y \,,        \\
\dot y &=& Bp_x + Cp_y \,,        \\
\dot p_x &=& - \frac{\partial V}{\partial x} \,, \\
\dot p_y &=& - \frac{\partial V}{\partial y} \,. 
\end{eqnarray}
This first order system of equations in $(x,y,p_x,p_y)$ can 
be easely recast in the
equivalent second order system of equations for $(x,y)$,
\begin{eqnarray}
\label{eqnova1}
\ddot x + \Omega^2 x &=& - A\frac{\partial V}{\partial x} 
- B\frac{\partial V}{\partial y} + \Omega^2 x \,, \\
\label{eqnova2}
\ddot y + \Omega^2 y &=& - B\frac{\partial V}{\partial x} 
- C\frac{\partial V}{\partial y} + \Omega^2 y \,, 
\end{eqnarray}
where the terms proportional to $\Omega^2$ were conveniently
added to both sides.
The comparison of equation (\ref{eqnova2}) with (\ref{erma2}) 
leads to the conclusion that the admissible frequencies must satisfy
\begin{equation}
\label{Omega}
\Omega^2 = \frac{1}{y}(B\frac{\partial V}{\partial x} + 
C\frac{\partial V}{\partial y})\,.
\end{equation}
Also, the comparison of equation (\ref{eqnova1}) with equation 
(\ref{erma1}), when $\Omega$ is given by (\ref{Omega}),
shows that the potential must obey the linear first 
order partial differential  equation
\begin{equation}
\label{eqv}
(Bx - Ay)\frac{\partial V}{\partial x} + (Cx - By)\frac{\partial
V}{\partial y} = \frac{1}{x^2}F(y/x)\,.
\end{equation}
The characteristic equations associated to (\ref{eqv}) can be
written in the form
\begin{equation}
\frac{dx}{Bx - Ay} = \frac{dy}{Cx - By} = \frac{x^{2}dV}{F(y/x)}\,,
\end{equation}
and have - as can be easely checked - the general solution
\begin{equation}
\label{V}
V = \frac{1}{2}\Lambda(q,t) + \frac{1}{q}\int^{s}F(s')ds'\,.
\end{equation}
Here $\Lambda(q,t)$ is an arbitrary function of its arguments, and 
the new variables $q$ and $s$ are defined in terms of the 
dynamic variables and the parameters of the Hamiltonian as 
\begin{eqnarray}
q &=& Ay^2 - 2Bxy + Cx^2 \,,\\
s &=& y/x \,.
\end{eqnarray}
It is also convenient to define the function
\begin{equation}
\xi(s) \equiv As^2 - 2Bs + C \,,
\end{equation}
so that $q = x^{2}\xi(s)$. 

In the conventional notation, 
the resulting Hamiltonian ELRR system implied by the admissible
frequencies and potentials, now reads 
\begin{eqnarray} 
\label{ema1} 
\ddot x + \rho\frac{\partial\Lambda}{\partial q}x &=& 
\frac{1}{yx^2}\bar{f}(y/x) \,,\\
\label{ema2} \ddot y + 
\rho\frac{\partial\Lambda}{\partial q}y &=& 
\frac{1}{xy^2}\bar{g}(x/y) \,, 
\end{eqnarray}
 where 
\begin{eqnarray}
\bar{f}(s) &=& 2\rho\frac{s}{\xi^2}\int^{s}F(s')ds' +
\frac{s}{\xi}(As - 
B)F(s)\,,  \\
\bar{g}(1/s) &=& 2\rho\frac{s^3}{\xi^2}\int^{s}F(s')ds' +
\frac{s^2}{\xi}(Bs - C)F(s)\,,  
\end{eqnarray}
and
\begin{equation}
\rho = AC - B^2 \,.
\end{equation}

There remains two arbitrary functions in the Hamiltonian ELRR
system, namely $\Lambda(q,t)$ and $F(y/x)$.
In fact, $\bar{f}$ and $\bar{g}$ in (\ref{ema1}--\ref{ema2}) are
determined by the single homogeneous function $F$, and
one can check directly that they satisfy
\begin{equation}
\label{constraint}
s(Bs - C)\frac{d\bar{f}}{ds} + (C + 2Bs)\bar{f} = (As -
B)\frac{d\bar{g}}{ds} +
(A + \frac{2B}{s})\bar{g} \,.
\end{equation}

It is interesting to  compare this result with those in the 
literature.  Cerver\'o and Lejarreta's  Hamiltonian 
Ermakov systems \cite{lejar} are obtained from the present 
formalism by setting $A = C = 1$, $B = 0$, and $\Lambda = 
\omega^{2}(t)q$ in (\ref{H}) and (\ref{V}). The resulting functions 
$\bar{f}$ and $\bar{g}$  do satisfy their Hamiltonian constraint 
 \cite{lejar}, which is precisely relation 
(\ref{constraint}) specialized for the appropriated parameter 
values. This formalism has already been used to study the 
propagation of elliptic gaussian beams in nonlinear, dispersive 
media \cite{goncha,nturo}. Also the completely integrable 
class of Ermakov systems determined by Goedert \cite{goed} is 
Hamiltonian and derivable from the present formalism.  In this 
case the right choice is $A = C = 0$, $B = 1$, and  $\Lambda = 
2\int^{-q/2}w^{2}(q')dq'$.  Needless to say, the functions $\bar{f}$ 
and $\bar{g}$ resulting from this prescription do satisfy the 
integrability condition stated in \cite{goed}.  

According to Liouville-Arnold theorem,  $2n-$dimensional 
Hamiltonian systems possessing $n$ independent constants of motion 
in involution with compact level surfaces are integrable by 
quadratures \cite{arnold}. For these systems, the motion is 
quasiperiodic and restricted to $n-$dimensional tori. The present 
class of four dimensional Hamiltonian ELRR systems will posses two 
independent constants of motion in involution provided that the 
Hamiltonian does not depend on time. In this case, $H$ itself 
is a constant of the motion independent of the LRRI. So, when the 
function $\Lambda(q,t)$, which is the source of time-dependence in 
the Hamiltonian, does not contain $t$, one can expect that the 
problem is completely integrable. In fact, the level surfaces 
$H=$const. and $I=$const. are not compact in general, but the 
differential equations nevertheless are reducible to quadratures.  
Consequently the Hamiltonian ELRR system treated here must be 
included in the non trivial class of solvable ELRR systems, to 
which belong, for instance, some systems analyzed by Govinder and 
Leach who considered frequency functions depending only on time 
\cite{gole}.  

The Hamiltonian formalism can be used to solve elegantly the 
equations of motion. Clearly $x$ and $y$ are not the natural 
coordinates for the problem. Changing from $(x,y)$ to the new
coordinates $(q,s)$, recasts the 
Hamiltonian and the LRRI in the form 
\begin{eqnarray}
\label{newH}
H &=& 2\rho q p_{q}^2 + \frac{1}{2}\Lambda(q,t) +
\frac{I(s,p_s)}{q}\,,\\
I &=& \frac{1}{2}\xi^{2}(s)p_{s}^2 + \int^{s}F(s')ds'\,,
\end{eqnarray}
where $p_q$ and $p_s$ are the momenta conjugate to $q$ and $s$ 
respectively.  This transformation decouples the dynamics and 
allows to treat separately the subsets $(p_{q},q)$ and 
$(p_{s},s)$ of the phase space.  Moreover, when $\Lambda$ is time-
independent, $H$ is also an invariant. In this situation, we can 
proceed in a manner similar to that used in the energy integral 
method of standard classical mechanics and split the problem in 
two separable ordinary differential equations, 
\begin{eqnarray}
\label{num1}
(dq/dt)^2 &=& 4\rho(2qH - q\Lambda(q) -2I)\,, \\
\label{num2}
(ds/dt)^2 &=& 2q^{-2}(t)\xi^{2}(s)\left(I -
\int^{s}F(s')ds'\right).
\end{eqnarray}
Equations (\ref{num1}-\ref{num2}) can be successively solved in 
terms of quadratures yielding a formal solution to the  ELRR 
system.  In fact, the solution of (\ref{num2}) requires the 
knowledge of $q(t)$ obtained from (\ref{num1}). 

The structure of (\ref{num2}) suggests the rescaling of the 
time variable according to
\begin{equation}
\label{ntime}
d\tau(t) = dt/q(t) \,.
\end{equation}
Such rescaling is applicable globally only when $q(t)$ is positive 
definite in time and $\tau$ results monotonic. 
Under this circuntances system (\ref{num1}-\ref{num2}) reads
\begin{eqnarray}
\label{numq}
\left(\frac{dq}{d\tau}\right)^2 &=& 4\rho{q^2}\left(2qH 
- q\Lambda(q) - 2I\right) \,, \\
\label{num3}
\left(\frac{ds}{d\tau}\right)^2 &=& 2\xi^{2}(s)\left(I -
\int^{s}F(s')ds'\right)\,,
\end{eqnarray}
which is a decoupled set 
of separable equations and, therefore, reducible to quadratures. 
This procedure specifies a {\it general} solution of the problem 
that involves four arbitrary constants, namely $H$, $\,I$ and two
constants arising from (\ref{numq}) and (\ref{num3})

One interesting remark concerning the Hamiltonian Ermakov system 
is the following: changing $\Lambda(q)$ according to $\Lambda(q) 
\longmapsto \Lambda(q) + k_1$ or according to $\Lambda(q) 
\longmapsto \Lambda(q) + 2k_2/q$, 
where $k_1$ and $k_2$ are constants, is equivalent to changing 
the values of the contants $H$ or $I$ in (\ref{num1}) and, 
therefore, will not change the nature of the integral to be 
performed.  This transformation can be explored to identify 
variations of a known, exactly solvable, systems that are also 
exactly solvable.  In one hand the addition of $k_1$ to $\Lambda$
does not lead to a relevant variation of the problem, since it 
only implies a change in the numerical value of $H$. On the other
hand, the addition of $2k_{2}/q$ to $\Lambda$ does imply a {\it 
qualitative} change in the potential with no relevant change in 
the calculations. We only need to replace $I$ in the 
original equations according to 
\begin{equation}
I \longmapsto I + k_2 \,.
\end{equation}

As a concluding remark to this section, we stress that the 
Hamiltonian ELRR has been reduced to quadratures. Whether these 
quadratures can actually be performed globally is a different 
question to be examined in each particular application. Another 
important follow-up remark concerns the nonlinear superposition 
law \cite{ray} associated to the ELRR systems.  In general, this 
nonlinear superposition law is implicit in the sense that it 
cannot be actually applied in view of the coupling between the 
equations.  However, when at least one of the equations decouples, 
the integration can be carried through and the corresponding 
nonlinear superposition law becomes explicit.  For Hamiltonian 
ELRR systems, we arrive at equations (\ref{ntime}) and (\ref{num3}) 
which constitute an explicit nonlinear superposition law. This is 
clear since $s(t)$ is constructed using $q(t)$ obtained from a 
decoupled equation, namely equation (\ref{num1}).

\section{Sample applications}
In this section we work out some sample applications of the 
theory.  In particular we analyze the Calogero potential   
and a super integrable example of a noncentral potential. 
Several other potentials which represent generalizations of either 
the harmonic or the Coulomb potentials could be treated in a 
similar way at least in what concerns the dynamics of the 
two-dimensional motion obtained by projection in an appropriate plane.  
The super integrable Hartmann potential and some generalizations 
or variations of it (see \cite{evans,boschi}) certainly belong to this 
category. A generalized version of the coupled Pinney 
equations of interest in two-layer shallow-water wave theory 
\cite{ath} can also be solved analitically by the formalism of 
this paper.

\subsection{The Calogero system as a Hamiltonian ELRR system}

As a first example to illustrates the analytical integration of a 
Hamiltonian ELRR system, we consider the Calogero potential and 
its
associated system \cite{Calo}, which is a one-dimensional three 
body problem given by the Hamiltonian 
$$
H_C = \frac{1}{2}(p_{1}^2+p_{2}^2 + p_{3}^2) +
\frac{\sigma^{2}}{6}((x_1 - x_2)^2 + (x_2 - x_3)^2 + (x_3 -
x_1)^{2}) 
$$
\begin{equation}
\label{HCal}
+\frac{g_1}{(x_1 - x_2)^2} + \frac{g_2}{(x_2 - x_3)^2} +
\frac{g_3}{(x_3 - x_1)^2} \,,
\end{equation}
where $\sigma, g_{1}, g_{2}$ and $g_3$ are non negative constants. 
A rescaling of time and space coordinates allows to set $\sigma\equiv 1$.
Moreover, in view of the translational invariance of the problem, we 
transform to the center of mass and Jacobi coordinates 
\begin{eqnarray}
\label{cmass}
R &=& \frac{1}{3}(x_1 + x_2 + x_3)\,, \nonumber \\
\label{jacobix}
x &=& \frac{1}{\sqrt{2}}(x_1 - x_2) \,, \\
\label{jacobiy}
y &=& \frac{1}{\sqrt{6}}(x_1 + x_2 - 2x_3) \,. \nonumber
\end{eqnarray}
The center of mass only executes free motion, and the $(x,y)$
dynamics is described by the reduced Hamiltonian
\begin{equation}
\label{HCE}
H = \frac{1}{2}(p_{x}^2 + p_{y}^2) + \frac{1}{2}(x^2 + y^2) +
\frac{g_1}{2x^2} + \frac{2g_2}{(x -\sqrt{3}y)^2} + \frac{2g_3}{(x
+ \sqrt{3}y)^2} \,.
\end{equation}
As it can be easily checked, the reduced Hamiltonian (\ref{HCE}) is of 
Ermakov type.  Moreover, being autonomous, it is integrable.  The 
coefficients $A, B$ and $C$ in (\ref{H}) become, in this case, 
$A =C= 1$ and $B = 0$. Consequently, $q = x^2 + y^2 = r^2$ and the 
corresponding Ermakov potential reads 
\begin{equation}
\label{VCE}
V = \frac{q}{2} + \frac{1 + s^2}{2q}\left(g_1 +
\frac{4g_2}{(1 - \sqrt{3}s)^2} + \frac{4g_3}{(1 +
\sqrt{3}s)^2}\right) \,.
\end{equation}
By comparing this form and equation (\ref{V}) we find that 
for the Calogero system,
\begin{equation}
\label{lll}
\Lambda = q \,,
\end{equation}
and
\begin{equation}
\label{fff}
\int^{s}F(s')ds' = \frac{1 + s^2}{2}\left(g_1 + \frac{4g_2}{(1 -
\sqrt{3}s)^2} + \frac{4g_3}{(1 + \sqrt{3}s)^2}\right) \,.
\end{equation}
Equation (\ref{num1}) can be solved analytically for 
$\Lambda$ given in (\ref{lll}) and it can be verified directly that 
the rescaling $t \mapsto \tau$ is properly defined, that is, $q(t)$ 
is positive definite in time.
The resulting system of equations to be solved are now
\begin{eqnarray}
\label{qqq}
\left(\frac{dq}{d\tau}\right)^2 &=& 4q^{2}(2Hq - q^2 - 2I) \,, \\
\label{sta}
\left(\frac{ds}{d\tau}\right)^2 &=& (1\!+\!s^2)^2\left(2I\!-\!(1\!+\!s^2)
\left(g_1\!+\!\frac{4g_2}{(1\!-\!\sqrt{3}s)^2} + 
\frac{4g_3}{(1\!+\!\sqrt{3}s)^2}\right)\right).
\end{eqnarray}

It is now easy to find the solution $q(\tau')$, where we make the 
convenient replacement $\tau\mapsto\tau'\equiv\sqrt{2I}\tau$ ($I$ is
strictly positive for the Calogero system):
\begin{equation}
\label{qtauprime} 
q(\tau^\prime)=\frac{2I}{H-\sqrt{H^2-
2I} \sin(2(\tau^\prime + c_1))}\, \,,
\end{equation}
where $c_1$ is the integration constant arising from (\ref{qqq}).
This integration constant and the integration constant $c_2$ arising
from the integration of (\ref{sta}) below can always be expressed
in terms of the initial position and of $H$ and $I$.

The function $s(\tau)$ can be evaluated in closed form for a few
different values of $g_i$. We choose $g_1 = g_2 = g_3 \equiv g$, 
which is perhaps the most interesting case yielding solutions
in terms of circular functions
\begin{equation}
\label{eqtauprime}
s(\tau^\prime) = \tan\left(\frac{1}{3}\sin^{-1}\left(\left(1 -
\frac{9g}{2I}\right)^{1/2}\sin(3(\tau^\prime + c_2))\right)\right) \,,
\end{equation}
where $c_2$ is the integration constant refered to before.

The last two equations express the parametric orbits of the 
problem and is a general solution involving 
four integration constants, namely 
$H, I, c_1$ and $c_2$.  
This solution is in full agreement with the results of Khandekar 
and Lawande \cite{Khan}, but are obtained in a more systematic way.  

As mentioned before, the addition of $2g_{4}/q$ to $\Lambda$,  
where $g_4$ is a constant,
does not alter the nature of the analytic solutions.  It only 
requires the change $I \rightarrow I + g_4$ in all formulae. This
proves the existence of closed form solutions for the modified 
Hamiltonian 
\begin{eqnarray}
H &=& \frac{1}{2}(p_{x}^2 + p_{y}^2) + \frac{\sigma^2}{2}(x^2 + y^2)
+\frac{g_1}{2x^2}
\nonumber \\
&+& \frac{2g_2}{(x -\sqrt{3}y)^2} + \frac{2g_3}{(x
+ \sqrt{3}y)^2} +  \frac{g_4}{x^2 + y^2} \,.
\end{eqnarray}
Therefore, in the original variables, the
one-dimensional three body problem described by the Hamiltonian 
$$
\bar{H}_C = \frac{1}{2}(p_{1}^2 + p_{2}^2 + p_{3}^2) +
\frac{\sigma^{2}}{6}\left((x_1 - x_2)^2 + (x_2 - x_3)^2 + (x_3 -
x_1)^{2}\right)+
$$\vskip -25pt
\begin{equation}
 \frac{g_1}{(x_1 \!-\! x_2)^2} +
\frac{g_2}{(x_2 \!-\! x_3)^2} +
\frac{g_3}{(x_3 \!-\! x_1)^2} 
+ \frac{3g_{4}}{(x_1 \!-\! x_2)^2 
+ (x_2 \!-\! x_3)^2 + (x_3 \!-\!x_1)^{2}} 
\end{equation}
is also exactly solvable.  $\bar{H}_C$ can be viewed as an 
integrable modification of the Calogero system. Notice that the 
term in $g_4$ is a new contribution that did not 
belong to the original system. 
We remark that this generalization is possible 
thanks to the fact that, in Jacobi coordinates, the Calogero 
system possesses the structure of a Hamiltonian ELRR system.  

\subsection{A Hamiltonian with dynamic symmetry}

The noncentral problem described by the Hamiltonian
\begin{equation}
\label{super}
H = \frac{1}{2}(p_{x}^2 + p_{y}^2) - \frac{\sigma}{\sqrt{x^2 +
y^2}} + \frac{g_1}{y^2} + \frac{g_{2}x}{y^{2}\sqrt{x^2 +
y^2}} \,, 
\end{equation}
where $\sigma$, $g_1$ and $g_2$ are positive constants, is known to 
possess a dynamic symmetry group \cite{winternitz} and is 
separable in parabolic and polar coordinates. 
Three dimensional 
extensions of $H$ are super integrable systems \cite{evans}, and 
have received attention as a noncentral force problem amenable to 
Feynman quantization \cite{boschi}. Several other two-dimensional 
versions of three dimensional super integrable models can be 
reduced to Hamiltonian ELRR systems. We selected (\ref{super}) as 
a good example to illustrate the technique proposed in section $2$.  

A rescaling of time and space allows to set $\sigma \equiv 2$
without any loss of generality.
For this problem also $q = x^2 + y^2 = r^2$ and it is convenient  
to introduce polar coordinates so that $s = \tan\theta$. The
potential, in $(r,\theta)$ coordinates, now reads
\begin{equation}
\label{eq45}
V = - \frac{2}{r} + 
\frac{1}{r^{2}\sin^{2}\theta}\left(g_1 + g_{2}\cos\theta\right) \,.
\end{equation}
Comparison of equations (\ref{eq45}) and (\ref{V}) shows that in 
this case
\begin{equation}
\Lambda(q) = - \frac{4}{r}  \,,
\end{equation}
and
\begin{equation}
\int^{s = \tan\theta}F(s')ds' = \frac{1}{\sin^{2}\theta}
\left(g_1 + g_{2}\cos\theta\right) \,. 
\end{equation}

Let us restrict considerations to the cases where the rescaled time $\tau$ is
well defined and monotonic. A detailed analysis shows that for positive 
definite LRRI invariant (sufficiently high angular momentum)
the trajectories never cross the
origin. For $I > 0$ the variable $q = r^2$  never vanishes and
$\tau$ is monotonically increasing  as can be calculated 
directly from equation (\ref{ntime}). We note that for $I>0$, $q\neq0$
is also required for the right
hand side of the equation (\ref{num1}) to be 
positive definite, a necessary condition for the
existence of real valued solutions. We therefore consider $I>0$ and
reduce the original problem to the set of differential equations
\begin{eqnarray}
\label{posi}
\left(\frac{dr}{d\tau}\right)^2 &=& 2r^{2}(Hr^2 + 2r - I) \,, \\
\label{positi}
\left(\frac{d\theta}{d\tau}\right)^2 &=& 2\left(I - \frac{1}{\sin^{2}\theta}
(g_1 + g_{2}\cos\theta)\right) \,.
\end{eqnarray}
A direct integration of (\ref{posi}--\ref{positi})
(with $\tau\mapsto\tau'\equiv
\sqrt{2I}\tau$) yields the parametric orbit equation
\begin{eqnarray}
r(\tau') &=& \frac{I}{1 - \sqrt{1 + HI}\sin(\tau' + c_1)} \,\,, \\
\label{teta}
\cos\theta(\tau') &=& -\frac{g_2}{2I}\left(1 - \sqrt{1 + 
4I(I - g_1)/g_{2}^2}\sin(\tau' + c_2)\right) \,, 
\end{eqnarray}
where $c_1$, $c_2$ are integration constants which can be 
expressed in terms of the initial position and of $H$ and $I$. 
For negative values of the energy, the motion is bounded. 
When the energy is positive we find open trajectories that
escape to infinity for finite values of $\tau'$. However in the
original parameter $t$ this process is regular and takes an infinite 
amount of time as expected on physical grounds. Another interesting
feature of the bounded motion is the fact that it does not explore
the entire range $\theta = 0$ to $\theta = 2\pi$. This is evident from
equation (\ref{teta}), which inplies $\chi_{-} \leq \cos\theta \leq
\chi_{+}$, where
\begin{equation}
\chi_{\mp} = - \frac{g_2}{2I}\left(1 \pm  \sqrt{1 +4I(I - g_1)/g_2^2}\right) \,.
\end{equation}
It is ease to verifie that $\chi_{+} < 1$ which 
corresponds to the fact that the trajectories never visit that
sector of the plane where $\theta \leq \arccos\chi_+$. 
It is also clear that for sufficiently small values of $g_2$
there may exist another excluded sector around $\theta = \pi$.
This is the case when $\chi_- > -1$ which is possible if and only if
$2I > g_2$ and $g_1 > g_2$.

As in the first example, the addition of a term $2g_{3}/q$ to 
$\Lambda(q)$, where $g_3$ is a new constant, does not alter the 
calculations. The new exactly solvable potential is given by 
\begin{equation}
V = \frac{1}{2}(p_{x}^2 + p_{y}^2) - \frac{\sigma}{\sqrt{x^2 + y^2}} +  
\frac{g_1}{y^2} + \frac{g_{2}x}{y^{2}\sqrt{x^2 + y^2}} + \frac{g_3}{x^2 + 
y^2} \,.
\end{equation}
Here we should stress that the term proportional do $g_3$ is novel 
and represent an incorporation to the original potential that  
preserves the integrability property. Again this was possible 
thanks to the Hamiltonian character of this Ermakov system.  

\section{Conclusions}

A quite general class of exactly integrable Hamiltonian ELRR 
system has been identified and solved.  The basic result comes 
from  the fact that an Ermakov system being Hamiltonian (with a 
quadratic form in the momenta) is exactly solvable in terms of 
quadratures. Also the ultimate equations for those systems are 
decoupled, a fact that leads to practical nonlinear superposition 
laws. 

Another important feature of the formalism stems from the 
fact that for each exactly solvable model there always exists a 
modified version of the problem that is also integrable analytically.  This 
provides a mechanism to spawn new integrable Ermakov systems.
Two examples of physically interesting applications were treated in 
detail to illustrate the practical value of the method.  The 
application scope of the method, however, is much wider and a 
detailed assessment of its reach is an open question that deserve 
further study.  

Some open questions are readily identified.  A first open question  
concerns the expansion of the method to higher dimensional Ermakov 
systems. From the point of view of Physics, this would be most 
interesting, mainly for three space dimensions.  A second open 
question concerns the Hamiltonian character of ELRR systems with 
velocity dependent frequency function.  Still another interesting 
question concerns the perturbations theory  of ELRR systems. This 
completely unexplored subject has not been touched so far, perhaps 
because of the lack of an Hamiltonian structure.  The results of 
this paper opens a new prospective for such issues.  

\bigskip\bigskip 
\leftline{\bf Acknowledgement} \smallskip 
This work was supported by the Brazilian Conselho Nacional de 
Desenvolvimento Cient\'{\i}fico e Tecnol\'{o}gico (CNPq) and 
Financiadora de Estudos e Projetos (FINEP).

\end{document}